\documentclass[preprint2]{aastex}
\def\z850{\mbox{z$_{850}$}}
\def\i814{\mbox{I$_{840}$}}
\def\dim#1{\mbox{\,#1}}
\def\hide#1{}

\begin{document}

\title{Star Formation in a Cosmological Simulation of Reionization}
\author{A.\ Gayler Harford\altaffilmark{1} and Nickolay Y.\
  Gnedin\altaffilmark{2,3,4}} 
\altaffiltext{1}{JILA, University of Colorado and National Institute of
Standards and Technology, Boulder, CO 80309, USA;
gharford@sif.colorado.edu}
\altaffiltext{2}{Particle Astrophysics Center, Fermi National
  Accelerator Laboratory, Batavia, IL  60510, USA; gnedin@fnal.gov} 
\altaffiltext{3}{Department of Astronomy \& Astrophysics, The
  University of Chicago, Chicago, IL 60637, USA}
\altaffiltext{4}{Kavli Institute for Cosmological Physics, The
  University of Chicago, Chicago, IL 60637, USA}

\begin{abstract}
We study the luminosity functions of high-redshift galaxies in
detailed hydrodynamic simulations of cosmic reionization, which are
designed to reproduce the evolution of the Lyman-$\alpha$ forest
between $z\sim5$ and $z\sim6$. We find that the luminosity functions and
total stellar mass densities are in agreement with observations when
plausible assumptions about reddening at $z\sim6$ are made. Our simulations 
support the conclusion that stars alone reionized the universe.
\end{abstract}


\section{Introduction}

An important issue in cosmology today is the timing and mechanism
of the reionization of the universe. Since the early work of
\citet{mhr99}, stars have been thought to be prime candidates 
for the energy source, since the Gunn-Peterson trough disappears
prior to the presumed epoch of quasars ($z>5$).

However, compounding the problem of identifying the primary sources of cosmic
reionization are recent HST observations of galaxies at $z\sim6$ 
\citep{bitb04,bsem04,yw04,gdfr04,hcck06,bibf06}. 
It remains controversial whether enough stellar luminosity has actually
been observed to account for reionization at $z\sim6$, as is seen
in the spectra of SDSS quasars \citep{wbfs03,fsbw06}.

On the other hand, both numerical simulations
\citep{g00a,hg03,sh03,g04} and semi-analytical models \citep{night06}
predict star formation rates at $z\sim6$ nearly as large as those at
$z\sim4$, in direct conflict with some of the observational studies
\citep{bsem04}, but in agreement with
recent studies of Lyman-$\alpha$ emitting galaxies at $z\sim6$
\citep{hcck06}, which suggest that the star formation rate at $z\sim6$ is
substantially higher than that found by HST observations.

While the observational determination
of the star formation rate at $z\sim6$ will most likely remain
controversial for some time, in this paper we focus on a theoretical
aspect of this problem. Of course, predicting {\it a priori\/} the
absolute value of the average star formation rate in the universe at
$z\sim6$ is not possible with our current understanding of the galaxy
formation process, in large part because the efficiency of star
formation in molecular clouds at high redshift is not known. However,
modern cosmological simulations are sufficiently robust to predict the
{\it ratio\/} of star formation rates at, say, $z\sim4$ and
$z\sim6$. Thus, we can legitimately attempt to answer the following
question:

\hide{
Theoretical predictions {\it a priori\/} of the
absolute value of the average star formation rate in the 
high redshift universe is difficult 
with our current understanding of the galaxy
formation process.  Much of the problem stems from
uncertainties in the efficiency of star
formation in molecular clouds at these redshifts.  Thus it is
with considerable interest that we have been following the
recent observational estimates of total stellar mass density of
these galaxies as observed in the infra-red.  These observations
provide a further constraint to challenge simulations.
Thus, we ask
}

{\it
  After imposing a requirement on a theoretical model to
  reproduce the observed evolution of the Lyman-$\alpha$ forest
  between $z=4$ and $z=6$, would it be possible, by appropriately
  adjusting the star formation efficiency as a free parameter, to fit  
  {\it simultaneously\/} galaxy luminosity functions at $z\approx4$
  and $z\approx6$?
}

In addition, we can use the recent observational estimates of the
total stellar mass density at $z\approx5$ as an extra constraint. 

In this paper we use cosmological simulations of reionization as
our theoretical model. The main advantage of our set of simulations is
that they are not only able to resolve high redshift galaxies, but, by
virtue of including radiative transfer of ionizing radiation, are also able
to reproduce the mean properties of the observed Lyman-$\alpha$ forest between
redshifts 5 and 6 \citep{gf06}. In particular, our simulations are
consistent with the end of the reionization at $z\approx6$, clearly
seen in the SDSS data \citep{fsbw06}.

Thus, we attempt to place both the observations of the Lyman-$\alpha$ forest
at $z>5$ and the observations of star-forming galaxies at the same
cosmic epoch into a unified picture, in which the same simulation is
required to fit {\it all\/} available observational data. 

We briefly describe the
simulations in \S 2, present the star formation rates in \S 3.1,
discuss numerical resolution issues in \S 3.2, present our
luminosity functions in \S 3.3, and discuss the accumulation of stellar
mass in \S 3.4.  In \S 4 we conclude and discuss the results.

\section{Method}

Simulations used in this paper have been run with the ``Softened
Lagrangian Hydrodynamics'' (SLH) code \citet{g00a,g04}. The
simulations include dark matter, gas, star formation, chemistry, and
ionization balance in the primordial plasma, and three-dimensional radiative
transfer. A flat $\Lambda$CDM cosmology with values of cosmological
parameters as determined by the first year WMAP data \citep{svp03} is
used throughout this paper.  

The radiative transfer is modeled using the Optically Thin Variable
Eddington Tensor (OTVET) method of \citet{ga01}. While OTVET is an
approximation, its main advantage is that it is self-consistently
coupled to the rest of the simulation code, and thus takes into
account possible feedbacks among star formation, radiative transfer,
and gas dynamics on spatial and temporal scales resolved in a simulation.

Star formation is incorporated in the simulations using a
phenomenological Schmidt law on scales below the resolution limit of
the simulation (i.e.\ stars are allowed to form wherever the gas 
sinks below the
resolution limit, irrespective of the large-scale environment or
mass of the dark matter halo they are forming within).  This law  
introduces a free parameter: the star formation efficiency
$\epsilon_{\rm SF}$ [as defined by eq.\ (1) of \citet{g00a}]. In order
to incorporate the uncertainty in this parameter in our results, we
use three different simulations in this paper. All of them have a 
simulation volume of $8h^{-1}$ comoving Mpc. The first two
simulations include $128^3$ dark matter particles and the same number
of quasi-Lagrangian mesh cells for the gas evolution. The third
simulation includes 8 times more ($256^3$) resolution elements in the
same simulation volume.

There are two adjustable parameters in our simulations: the efficiency
of star formation $\epsilon_{\rm SF}$ and the ionizing efficiency,
i.e.\ the amount of ionizing radiation 
emitted per unit mass of stars. This latter value depends upon the
fraction of ionizing photons escaping from the vicinity of a star
forming region to the spatial scales resolved in the simulation, a
value that is unknown and possibly variable. Our simulations make no
assumptons about the value of the ionizing efficiency\footnote{Note that
  the ionizing efficiency parameter is related to the fraction of
  photons escaped from the resolution limit of a simulation. It is
  {\it not\/} directly related to the escape fraction from the virial
  radius, a quantity often used in analytical models. Computing the
  escape fraction from the virial radius is a formidable computational
  task, requiring much higher resolutuon simulations that we use
  here.}; rather the ionizing efficiency is adjusted to fit the
observational data on the Gunn-Peterson absorption in the spectra of
SDSS quasars, as explained in \citet{gf06}.

The ionization history of the universe depends almost completely upon a
product of the star formation efficiency and the ionizing efficiency,
i.e.\ the factor that determines how many ionizing photons are emitted
per unit mass of dense and rapidly cooling gas (which is assumed to be
transforming into stars). For each run the ionizing efficiency is
selected so that the simulation is consistent with the observed
evolution of the mean transmitted flux in the Lyman-$\alpha$ forest at
$z\sim6$ \citep{wbfs03,s04,fsbw06}. The star formation efficiency is
then treated as the sole free parameter.

\begin{table}[t]
\caption{Simulation Parameters\label{sim}}
\begin{tabular}{lccccc}
\tableline
Run &
$\Delta x$\tablenotemark{a} &
$\Delta M$\tablenotemark{b} &
$\epsilon_{\rm SF}$\tablenotemark{c} &
$z_{f}\tablenotemark{d}$ \\
\tableline
\tableline
LowResHighSFR & 2    & $2.6\times10^7$ & 0.15 & 4 \\
LowResLowSFR  & 2    & $2.6\times10^7$ & 0.05 & 4 \\
HighResLowSFR & 0.64 & $3.2\times10^6$ & 0.05 & 5.1 \\
\tableline
\end{tabular}
\tablenotetext{a}{Spatial resolution in $h^{-1}\dim{kpc}$.}
\tablenotetext{b}{Mass resolution in $M_\odot$.}
\tablenotetext{c}{Star formation efficiency.}
\tablenotetext{d}{Final redshift of simulation.}
\end{table}
The parameters of the three simulations are listed in Table
\ref{sim}. The two smaller boxes (we label them ``LowRes'') are
identical, except for the value of the star formation efficiency,
which is different by a factor of 3 between the two simulations. The
larger simulation (``HighRes'') has the same star formation efficiency
as the second LowRes simulation. To make the referral to a specific
simulation clear, we use the names that reflect the resolution and
star formation efficiencies of our simulations.

Galaxies in the simulation are associated with gravitationally bound objects,
identified with the DENMAX algorithm \citep{bg91}.

Population synthesis is carried out using the STARBURST99
package \citep{lsgd99} with the high mass loss Geneva tracks 
as described previously \citep{hg03} using a 
metallicity of 0.001 (0.05 solar) and a galaxy escape fraction for
ionizing photons of 0.1. The results were adjusted to reflect a
three exponent initial mass function, the ``canonical'' function given
in \citet{weidner06}. Also included is the emission of extra
Lyman-$\alpha$ photons due to the recombination of ionized hydrogen. 
Throughout the paper AB magnitudes \citep{ok83} are
used exclusively. 

Reddening was computed according to \citet{calzetti97}. 
A color excess of 0.15 was included for $z\approx4$ galaxies
\citep{ssad01}. For galaxies at  $z\approx6$ we present results both
without reddening \citep{smb05} and with a color excess of 0.05
\citep{bibf06}.  For the luminosity functions in Figure \ref{mags_lf_fig}
unreddened luminosities were multiplied by a factor of 1.8 to make
them comparable to reddening with a color excess of 0.05.

\section{Results}

\subsection{Star Formation Rate in the Simulations}

\begin{figure}[t]
\epsscale{1.0}
\plotone{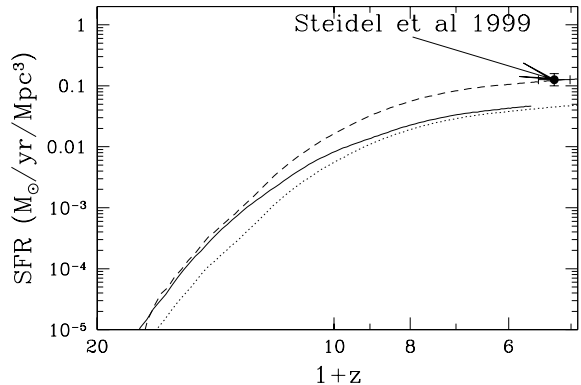}
\caption{Star formation rate as a function of redshift for our three
  simulations: HighResLowSFR (solid line), LowResLowSFR
  (dotted line), and LowResHighSFR (dashed
  line). The filled open circle shows the estimate of the average star
  formation rate at $z=4$ from \protect{\citet{sagd99}}. }
\label{sfr_fig}
\end{figure}
Figure \ref{sfr_fig} shows the evolution of the star formation history
in the three simulations described above. As we emphasized above, all
three runs are adjusted to reproduce the observed evolution of the
mean transmitted flux in the Lyman-$\alpha$ transition of neutral
hydrogen above $z\approx5$ reasonably well, so the reionization
histories are very similar in all three simulations and are consistent
with the existing data.
The star formation histories in the three runs are,
however, quite different. The two low resolution simulations differ by
approximately a factor of 3 by construction, since the star formation
efficiency is different by that factor between them. The high resolution
simulation has the same star formation efficiency as the
LowResLowSFR run, but forms more stars at earlier times
due to higher spatial resolution.
In all simulations the star formation rate begins to flatten
significantly at $z\la7$.

We have examined in detail two snapshots in time:  $z=4$
($1.5\dim{Gyr}$ after the Big Bang), the latest time of the low
resolution simulations, 
and $z=5.8$ ($0.99\dim{Gyr}$ after the Big Bang). Hereafter, unless specified
otherwise, we use our HighResLowSFR simulation as a fiducial model for the
$z=5.8$ snapshot, and the LowResLowSFR simulation as a fiducial model for
the $z=4$ snapshot; the HighResLowSFR simulation was not continued
until $z=4$ due to computational expense.

\begin{figure}[th]
\epsscale{1.0}
\plotone{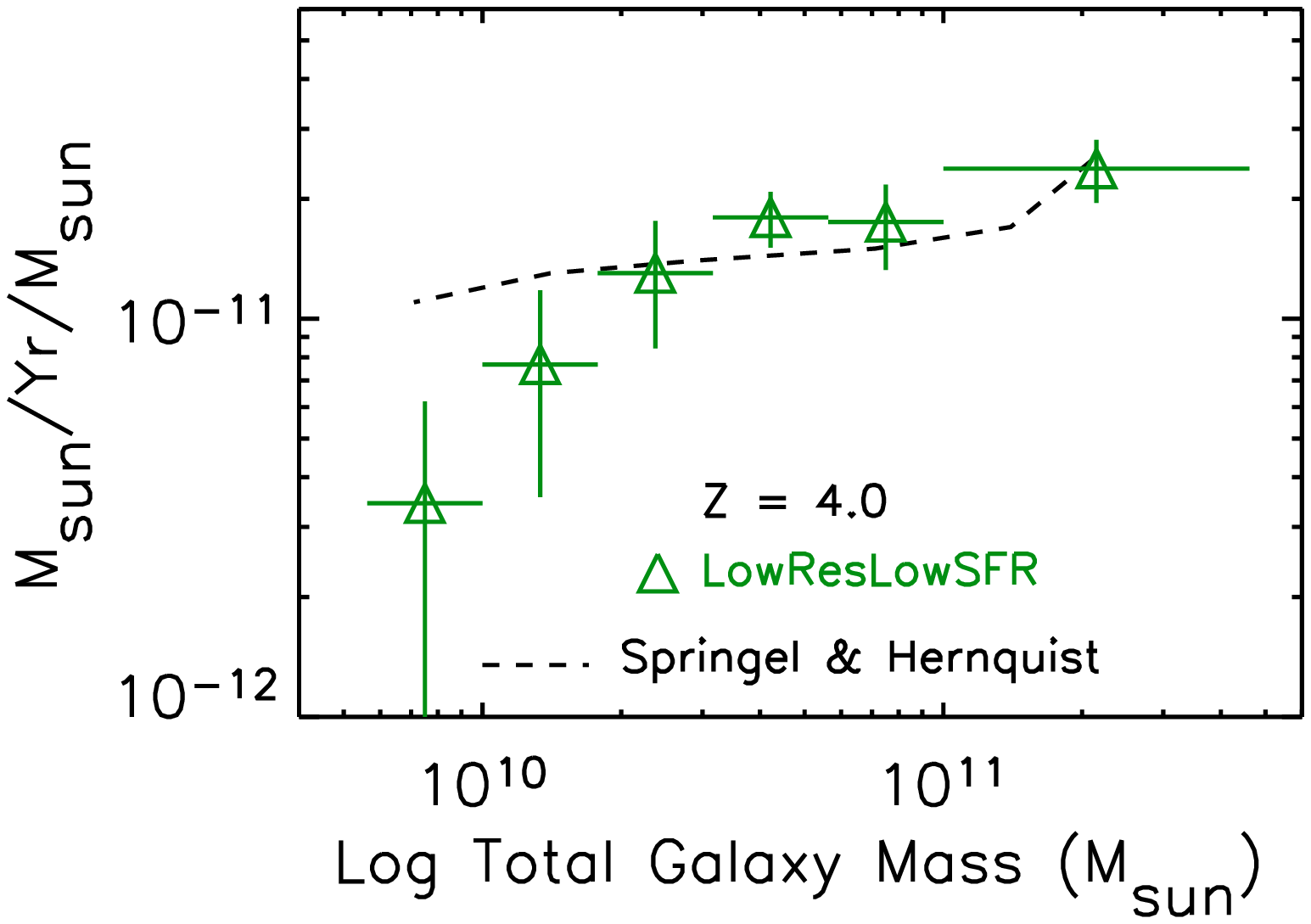}
\plotone{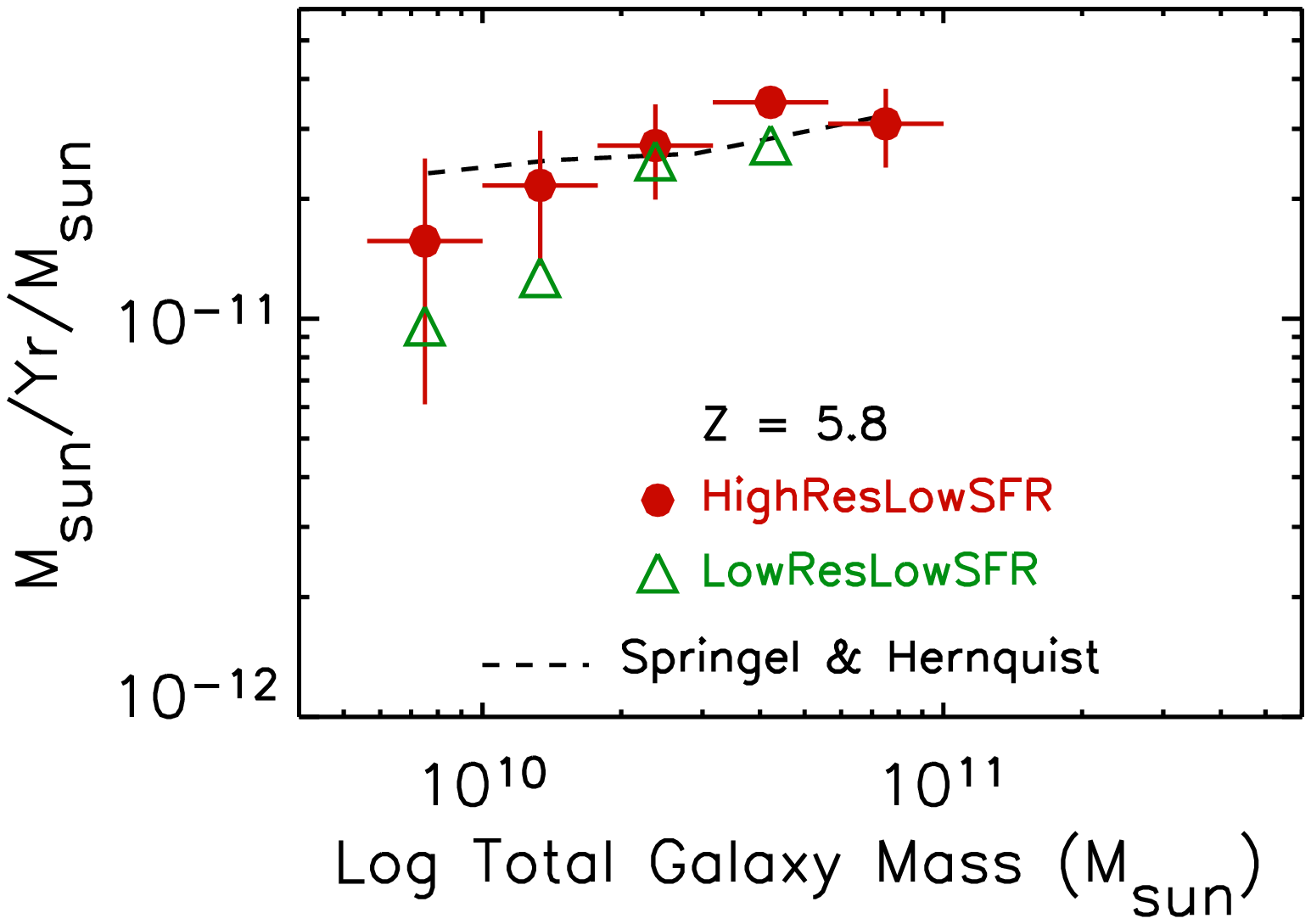}
\caption{Specific star formation rate (SFR per galaxy mass) vs the
  total galaxy mass for simulated galaxies at $z=4$ (top panel) and at
  $z=5.8$ (bottom panel). The horizontal lines represent galaxy mass
  ranges, and the vertical lines are standard deviations.  The lines
  are omitted for the LowResLSFR case in the bottom panel for clarity.
  The dashed lines are the results of \citet{sh03}.
}
\label{sfr_gal_fig}
\end{figure}

Deducing the global star formation rate by observing individual galaxies
requires knowledge of how star formation is distributed among galaxies
of different magnitudes. In order to test the robustness of the
simulations in predicting this dependence theoretically, we compare
our star formation rates in galaxies of various masses with the
simulations by \citet{sh03}, who conducted extensive resolution and
numerical convergence studies. The results of this comparison are shown
in Figure \ref{sfr_gal_fig}. As one can see, the LowResLowSFR run
underestimates specific star formation rates in galaxies with
$M\la2\times10^{10}\dim{M}_\odot$ at both redshifts, while the HighResLowSFR
is consistent with the results of \citet{sh03}.

It is important to underscore that the physics of gas cooling included in
our and \citet{sh03} simulations is rather different: while
\citet{sh03} only include equilibrium cooling of primeval (i.e.\
metal-free) plasma, in our simulations non-equilibrium radiative
transfer effects are taken into account. In addition, we include
cooling from heavy elements and molecular hydrogen. It is therefore
somewhat unexpected that the two sets of simulations agree so well. On
the other hand, if this agreement is not a mere coincidence, it
emphasizes the robustness of modern cosmological simulations in
predicting total rates of star formation in well resolved
galaxies (within the adopted phenomenological recipe of star
formation, of course). 

It is interesting to note that for the same mass the higher 
redshift galaxies have a somewhat greater star
formation rate than the lower redshift ones.  Nevertheless the total
star formation rate for these galaxies is greater at  $z=4$ than at $z=5.8$
because of the largest galaxies that have no counterpart at the earlier
redshift.  Since galaxies are observed only to limiting magnitudes,
the mass evolution of galaxies by itself helps to explain why the
luminosity function appears to evolve most at the bright end \citep{bibf06}.

The star formation rates shown in Fig.\ \ref{sfr_gal_fig} are measured
in the simulations as the amount of stellar mass added in the previous
$5\dim{Myr}$. We have verified that, at $z=6$, if that time interval
is extended 
up to $60\dim{Myr}$, the results barely change. Thus, galaxies in the
simulations have extended periods of star formation. This point is
important because it suggests that the observed luminous {\it UV\/} galaxies
which have been used to estimate the total stellar density are a
representative sample of all star-forming galaxies at $z\sim6$ 
\citep{stark06,yan06}.

\subsection{Effects of Numerical Resolution}

Comparisons of apples and oranges are rarely useful, so in order to
compare luminosities of simulated and observed galaxies, we need to know
whether observations count all the light emitted from individual
resolved galaxies, or if a substantial fraction of light remains
undetected below the sky surface brightness. If the former, then 
a comparison between the simulations and the
data is straightforward since it is easy to count all the emitted
light in the simulations.

If, however, a substantial fraction of the light is not detected in
the observations, the comparison with the simulations becomes much
more difficult. In this case a simulation is required not only to get
the total luminosity of the galaxy right, but also to correctly model
the full spatial distribution of this luminosity as well.
\footnote{One may argue that to get the total luminosity right, a
  simulation has to resolve the star forming regions as well. This is
  not the case, however, since all star formation prescription used in
  modern simulations incorporate free adjustable parameters, and one
  can phenomenologically fit the luminosity function of galaxies by
  properly adjusting parameters, without actually resolving the
  details of star formation.}
The latter presents a much more serious challenge to modern
simulations.

Unfortunately, we really have no way of knowing how much light is
missed in the observations. One possible way of approaching this
problem could be by using sufficiently high resolution simulations to
resolve star forming regions and recover the correct light
profile. We, therefore, first need to discuss the issues of numerical
resolution in our simulations, because in
our simulations galaxies do hide a significant albeit not dominant
fraction of their luminosity below the surface brightness limit of
current observations. Thus, we need to understand whether this is
merely a resolution effect, or a property real galaxies might have too.

Numerical resolution affects simulated galaxies in diverse and
complicated ways. Typically, simulations have constant spatial resolution in
comoving units, with the real space physical resolution deteriorating
with time. This effect would make later galaxies less concentrated
than earlier ones. For our simulations, this effect is not
tremendously large, since the difference between redshifts 4 and 5.8 is
only about 40\%. In addition, the cooling consistency
condition implemented in the SLH code \citep{g97} mitigates this
effect further. Therefore, we might expect that if our spatial
resolution is sufficient at $z=5.8$, then it is also sufficient at $z=4$.

However, one must also consider the mass resolution, whose effect usually
goes in the opposite direction. As time in the simulation progresses,
galaxies become more massive on average, i.e.\ they are represented by
a larger number of resolution elements (dark matter and stellar
particles, and gas cells). The number of resolution elements in a
given galaxy, if not sufficient, would affect the sharpness of
the central peak, which in turn could affect the spatial
distribution of star formation. In our fiducial HighResLowSFR
simulation a $10^{11}\dim{M}_\odot$ halo consists of about 30{,}000
dark matter particles, a comparable number of baryonic cells, and about
10{,}000 stellar particles. Is this number sufficient to localize
the star formation to within a few percent of the virial
radius?\footnote{For reference, the comoving virial radius of a
  $10^{11}\dim{M}_\odot$ halo is about $100$ comoving kpc, which
  translates into the angular diameter of $8^{\prime\prime}$ at $z=4$ and
  $7^{\prime\prime}$ at $z=5.8$.}

\begin{figure}[t]
\epsscale{1.0}
\plotone{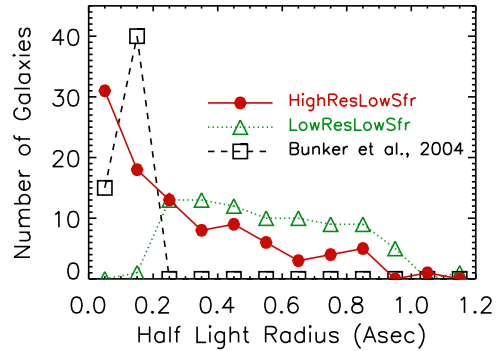}
\caption{Distributions of half light radii of simulated
  galaxies brighter than $\z850 = 29.5$ (unreddened)
  at $z\approx6$
  compared to observed galaxies \citep{bsem04}. For
  our cosmology $1\dim{arcsec}$ is 5.94 proper kpc.}
\label{rh_hist_fig}
\end{figure}

\begin{figure}[!ht]
\epsscale{1.0}
\plotone{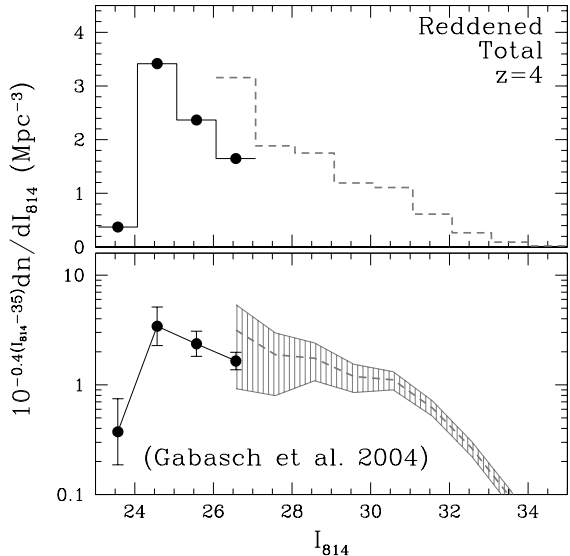}
\caption{Total reddened luminosity functions at $z=4$ for the
  LowResLowSFR simulation (dashed line) based upon galaxies
  larger than $10^{10}\dim{M}_\odot$.
  The solid black line with error
  bars shows the observational data from \citet{gbsh04}. The top
  panel shows the linear scale along the y-axis, while the bottom
  panel shows the log scale. The hatched band around the LowResLowSFR
  simulation shows Poisson errors.} 
\label{lf4_fig}
\end{figure}

\begin{figure}[!ht]
\epsscale{1.0}
\plotone{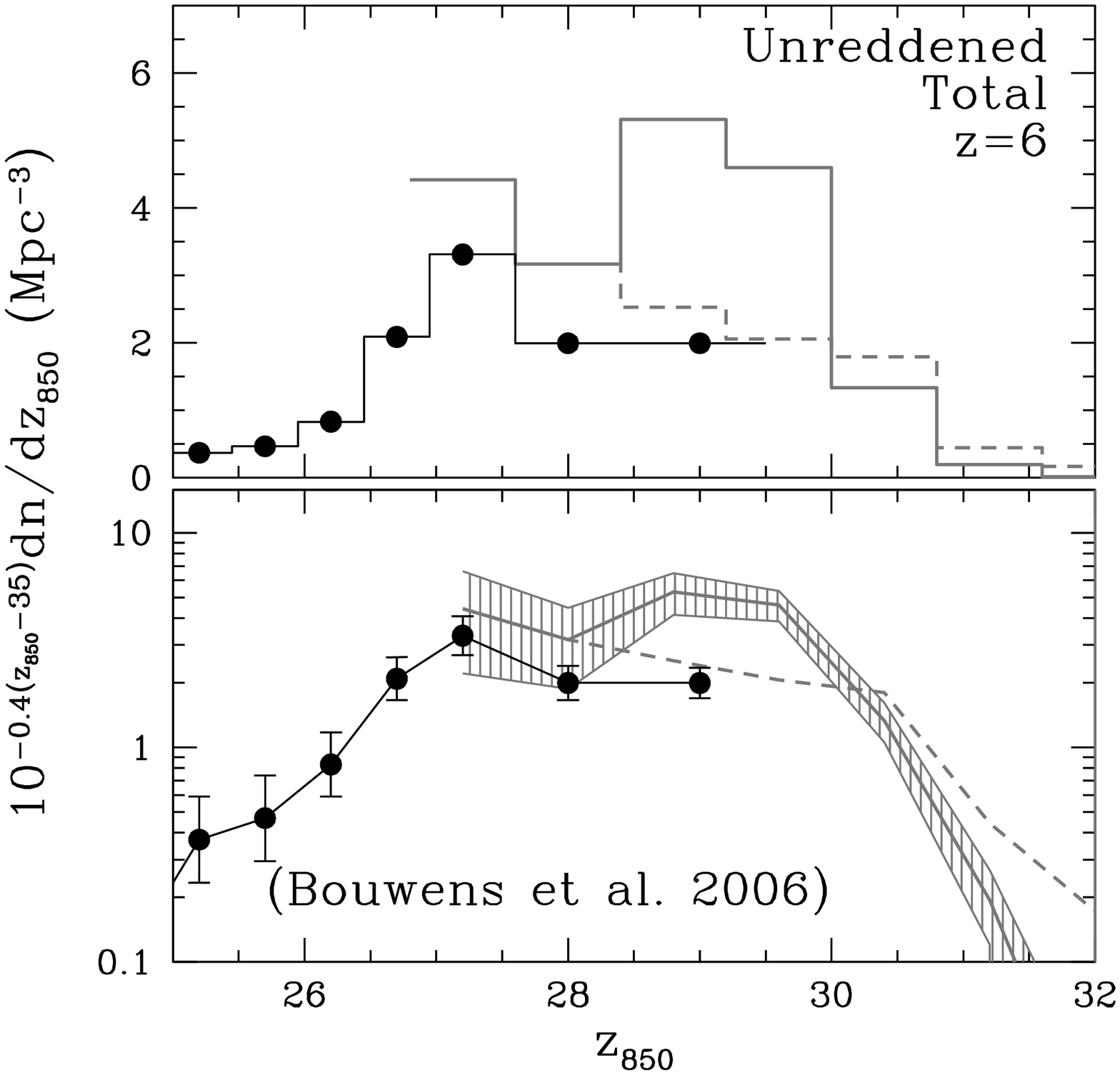}
\plotone{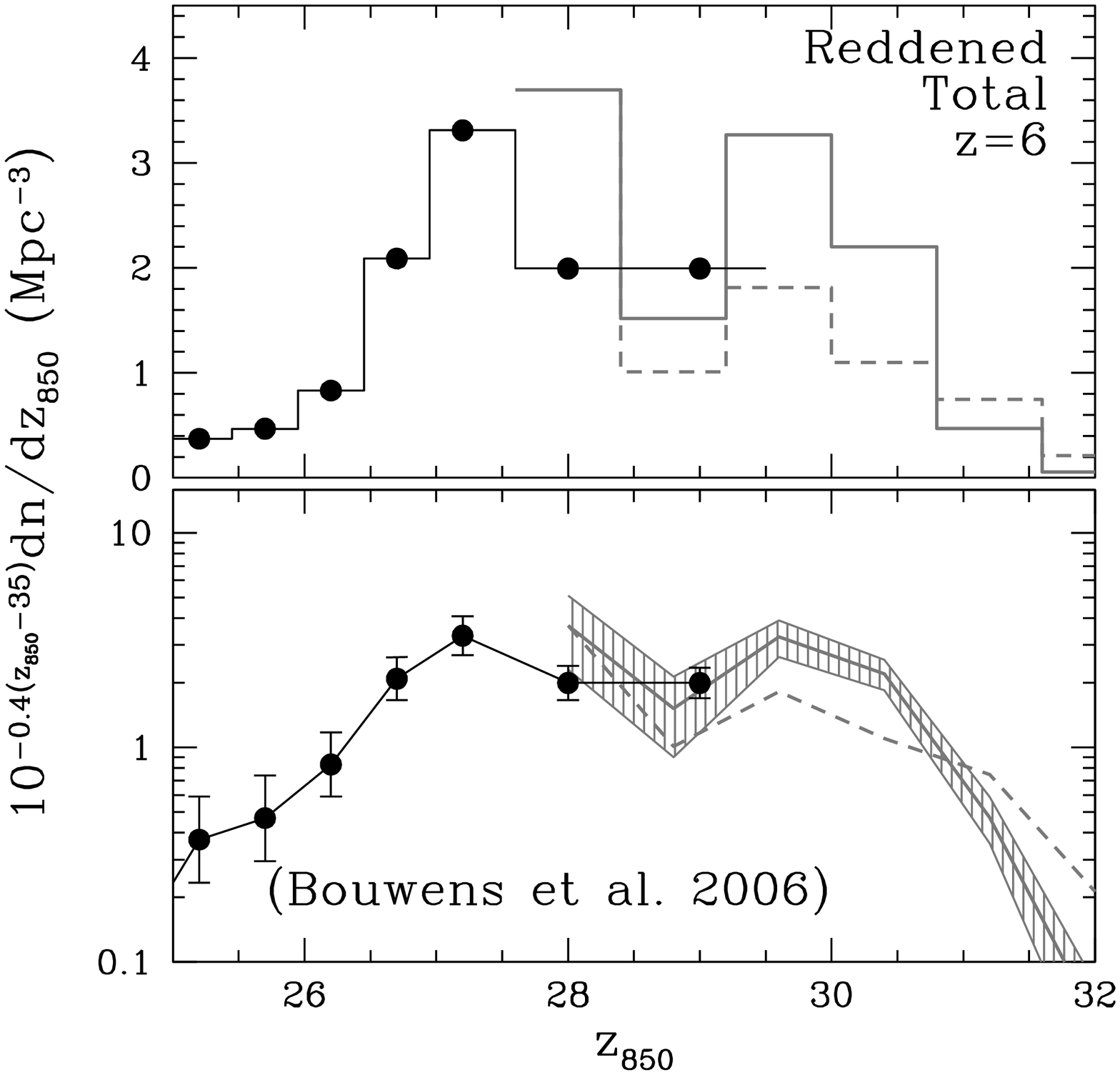}
\caption{The same as in Fig.\ \protect{\ref{lf4_fig}}, but for $z=5.8$,
  and showing in addition the HighResLowSFR run as the solid line within the
  hatched region. Two
  panels shows the unreddened and reddened luminosity functions
  respectively. The observational data are from \citet{bibf06}.} 
\label{lf6_fig}
\end{figure}

\begin{figure}[t]
\epsscale{1.0}
\plotone{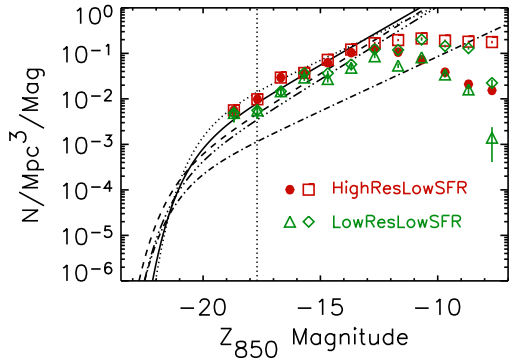}
\caption{Comparison of faint ends of luminosity functions at $z = 6$. 
Shown are luminosity
functions based upon the total galaxy population in the HighResLowSFR
(red) and LowResLowSFR (green) simulations. Filled circles and triangles 
include all galaxies with total mass $M\ga1.28\times10^{9}\dim{M}_\odot$ 
(about 100 particles), and squares and diamonds all galaxies with total mass
$M\ga3.84\times10^{8}\dim{M}_\odot$ (about 30 particles).
Since the deepest
observations extend only to the vertical dotted line, we have compared
our results to the Schechter function fits to the
observational data as presented in Table 12 of \citet{bibf06}. 
The line styles are
long dash \citep{bibf06}, dots \citep{dickin04}, 
short dash \citep{yw04}, dash dot \citep{bsem04}, 
and dash dot dot \citep{mal05}. The luminosity functions are independent
of the mass cut off out to at least $\z850<-15$. }
\label{mags_lf_fig}
\end{figure}

\begin{figure}[t]
\epsscale{1.0}
\plotone{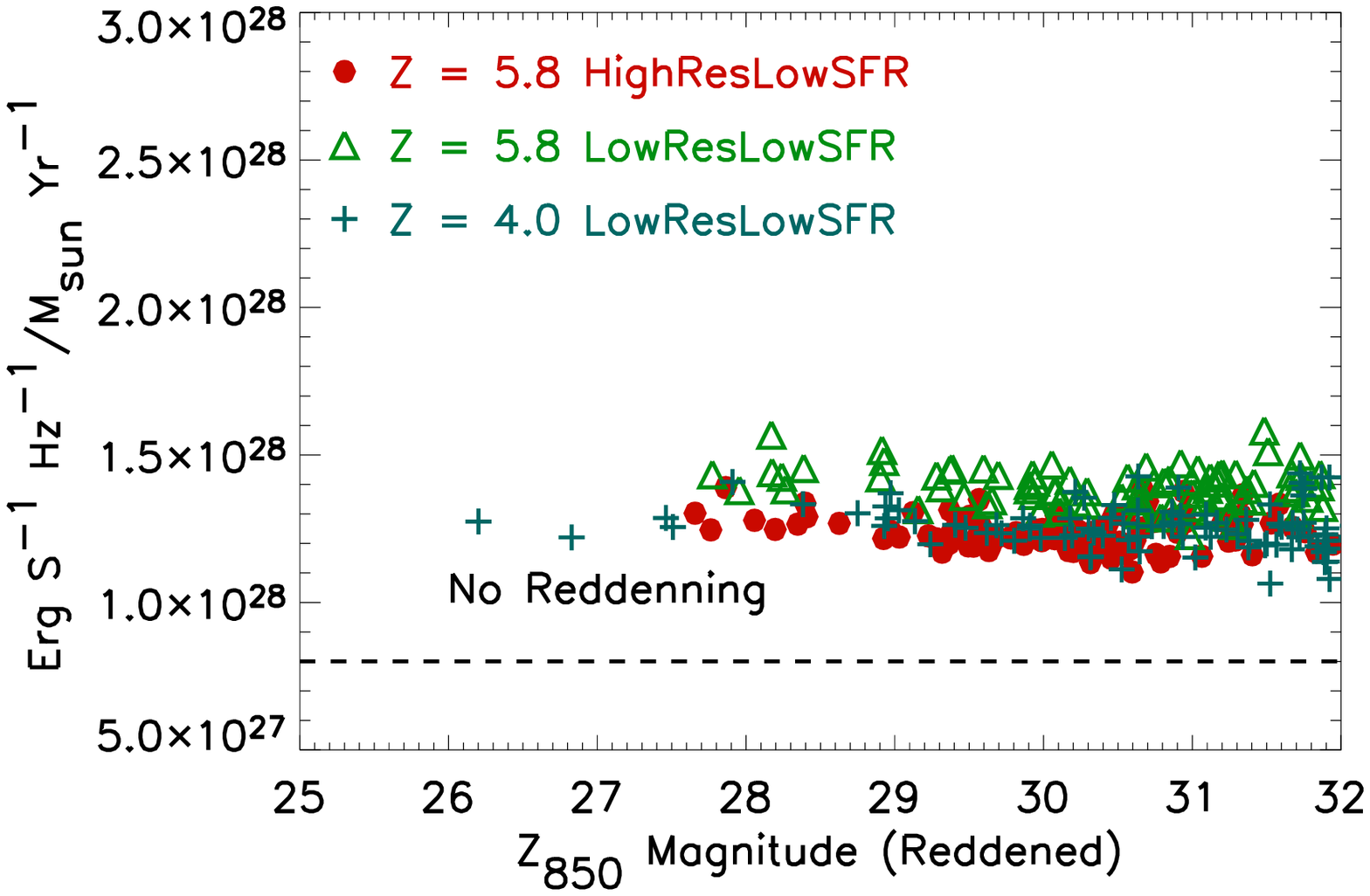}
\caption{ The ratio of star formation rate to the 
  stellar luminosity at $1500\dim{A}$ (as measured by the $\z850$ magnitude) 
  as a function of magnitude
  for the simulated galaxies. The luminosities are calculated from the
  unreddened magnitudes and plotted as a function of reddened
  magnitudes. Shown are galaxies
  larger than $10^{10}\dim{M}_\odot$.  The dashed line shows 
  the generally assumed value for this ratio \citep{mpd98}. }
\label{conv_fig}
\end{figure}

Figure \ref{rh_hist_fig} attempts to address this question by showing
the distribution of half light radii for simulated galaxies brighter
than $\z850 = 29.5$ (unreddened). The high resolution 
galaxies are more compact 
than the low resolution ones and, in some cases, as compact as
the observed ones, but there is still a substantial fraction (about 50\%)
of galaxies that are larger than the observed ones. Also, the
significant difference between the LowResLowSFR and HighResLowSFR runs
indicates that we have {\it not\/} reached the numerical convergence on the
half light radii of simulated galaxies.

Therefore, in the rest of this paper, we will assume that our
simulations do not properly resolve light profiles of most massive
galaxies, and we will use the total luminosities of model
galaxies.

\subsection{Luminosity Functions of Model Galaxies}

It is customary to present the luminosity function as a number density
of galaxies per unit luminosity, $dn/dL$, or the number of galaxies
per unit magnitude, $dn/dm$. In this paper we, however, are primarily
concerned with estimating the star formation rate 
at various redshifts from
luminosity functions, so the quantity of interest to us is the amount
of light emitted per unit log in luminosity $L^2 dn/dL$, or, 
equivalently, per unit magnitude $10^{-0.4m} dn/dm$. 

Figures \ref{lf4_fig} and \ref{lf6_fig} present the main result of
this paper. The total (i.e.\ including all the light of model galaxies
irrespective of surface brightness) luminosity functions 
from all our simulations are 
compared with the observational data from \citep{gbsh04} and \citep{bibf06}. 
We show luminosity functions both with linear and log vertical scale,
because the log plot allows one to see the whole range spanned by the values,
while the linear scale is useful because the total luminosity (and,
after an appropriate correction, the total star formation rate) is
simply an area under the curve, which can be
easily measured.

The overlap in magnitudes between the simulations and the observations
is not large; simulations, because of their limited box size, are not
able to produce the brightest, most rare galaxies. Observations, on the
other hand, are missing the majority of galaxies, that fall below
the flux limit. 

The advantage of using our choice for representing the galaxy
luminosity function can now be illustrated. For example, it does
appear that both the LowResLowSFR run and the observations are tracing the
same luminosity function at $z=4$. If we consider the combined curve
as a true luminosity function, it can be estimated from the top panel
of Fig.\ \ref{lf4_fig} that observations (without correcting for
incompleteness) account for about 50\% of the total star formation
rate, while the simulation includes about 60\% of the total SFR (with
the overlap between the simulation and observations being 10\% 
of the total). 

The LowResHighSFR simulation (not shown) is a factor of 
3 higher than the observational data,
which is not surprising, since it has a 3 times higher star formation
efficiency.  Unless specifically noted, we use the LowSFR simulations 
for comparisons to the observational data. 

The situation is different at $z=5.8$. In the interval
$27\la\z850\la29$, where the HighResLowSFR simulation overlaps with
the observational data from \citet{bibf06}, the simulation predicts a
factor of 2 more luminosity than the data if no reddening
correction is included, as argued by \citet{smb05}, which is about a
$3\sigma$ deviation. With the reddening correction of \citet{bibf06},
the simulation agrees with the data (accounting for about 70\% of the
total luminosity, while the data also account for about 70\% of the total
light). The HighResLowSFR simulation is approximately 
consistent with the LowResLowSFR run, indicating that our simulated
luminosity functions are close to the converged result (for the total
luminosities of modeled galaxies). 

We, thus, underscore the crucial importance of knowing the reddening
corrections at $z\sim6$ with sufficient precision.  Studies of reddening
agree that it is significantly less than at lower redshifts (see 
\citet{bibf06} for summary).

A matter of considerable interest and debate is the slope of the
faint end of the luminosity function when fitted to a Schechter function.
Figure \ref{mags_lf_fig} compares our luminosity functions
to various fits to observational data
from the literature.  The vertical line indicates the faintest
level observed, which is achieved in the recent study by \citet{bibf06}.
We refer the reader to this paper (Fig.\ 15 and accompanying
discussion)  for a comprehensive
comparison of luminosity functions at $z\approx6$ from the literature.

Simulation of the faint end of the luminosity function is limited by
the minimum size galaxy that can be adequately resolved by the simulation.
Since the  exact number of dark matter and gas particles needed is uncertain,
we show results for lower cutoffs of 100 and 30 particles for the LowResLowSFR
simulation.  To meaningfully compare the HighResLowSFR run we use the 
corresponding mass cutoffs (these will have 8 times as many particles).
The results are the same out to $\z850<-13$.  We find that, in contrast to
the large galaxies of the previous figures, the smaller galaxies tend to be
considerably less efficient at star formation.  For this reason, we think it
unlikely that a simulation with greater mass resolution would change the
faint end slope in the region brighter than $\z850<-15$, fainter than can be
currently observed.

\begin{figure}[t]
\epsscale{1.0}
\plotone{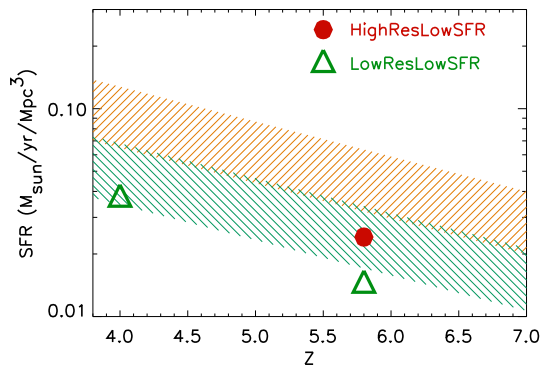}
\caption{The star formation history in simulated galaxies brighter than
  29.5 magnitude corrected for the finite size of the simulation box.
  The upper orange, hatched band shows the estimated range from
  \citet{bibf06} using the conventional conversion factor \citep{mfdg96}.
  To facilitate comparison with the simulated galaxies, which show
  a different conversion factor, the upper band
  has been transposed down by a factor of 1.5 (lower green band) as
  explained in the text.}   
\label{sfh_fig}
\end{figure}

To convert from luminosity to star formation rate, it is usually
assumed that a luminosity at $1500\dim{A}$ of
$8\times10^{27}\dim{erg/s/Hz}$ corresponds to 1 solar mass of newly
formed stars per year \citep{mpd98}.  This value, of course, depends
upon the assumptions one uses for the population synthesis.
We have used the more current ``canonical''
initial mass function given in \citet{weidner06}.  In addition,
we use a metallicity of $Z = 0.001$ (1/50th solar), which
agrees better with the simulation 
than does the usual solar metallicity
assumption (most of the simulated galaxies have metallicities 
well below 10\% solar.).   Figure \ref{conv_fig} shows that 
our conversion factors
are about 1.5 times higher than the conventional one.
Of course, we do not know the true
initial mass function and metallicity at these high redshifts.
We emphasize the difference here to facilitate comparison
of our results with the literature.

Figure \ref{sfh_fig}, which summarizes our findings, is based upon
a similar figure from \citet{bibf06}.  The values are given to a limiting
magnitude of 29.5.  The upper hatched region shows the original
range computed using the conventional conversion factor of
$8\times10^{27}\dim{erg/s/Hz}$ per $1M_\sun/\dim{yr}$.  To
facilitate comparison with our results, we also show this region transposed
downwards by a factor of 1.5 to form the lower hatched region, to
correspond to the conversion factor of
$1.2\times10^{28}\dim{erg/s/Hz}$ per $1M_\sun/\dim{yr}$ from Fig.\
\ref{conv_fig}. 
The star formation rates from our two LowSFR simulations {\it with the
  same magnitude cut\/} are also
shown as individual symbols at $z=4$ and $z=5.8$.
Both estimates
agree with the data, although our value is slightly lower at $z=4$
than the \citet{bibf06} estimate. The small discrepancy is due to the
fact that \citet{bibf06} estimate is based on the \citet{sagd99}
measurement, while our simulations agree best with \citet{gbsh04}, who
find a somewhat lower value of star formation rate at $z\sim4$.
\citet{gssb04} considers their results to be in accord with those of
\citet{sagd99}.

\subsection{Total Accumulated Stellar Mass Density}

\begin{figure}[t]
\epsscale{1.0}
\plotone{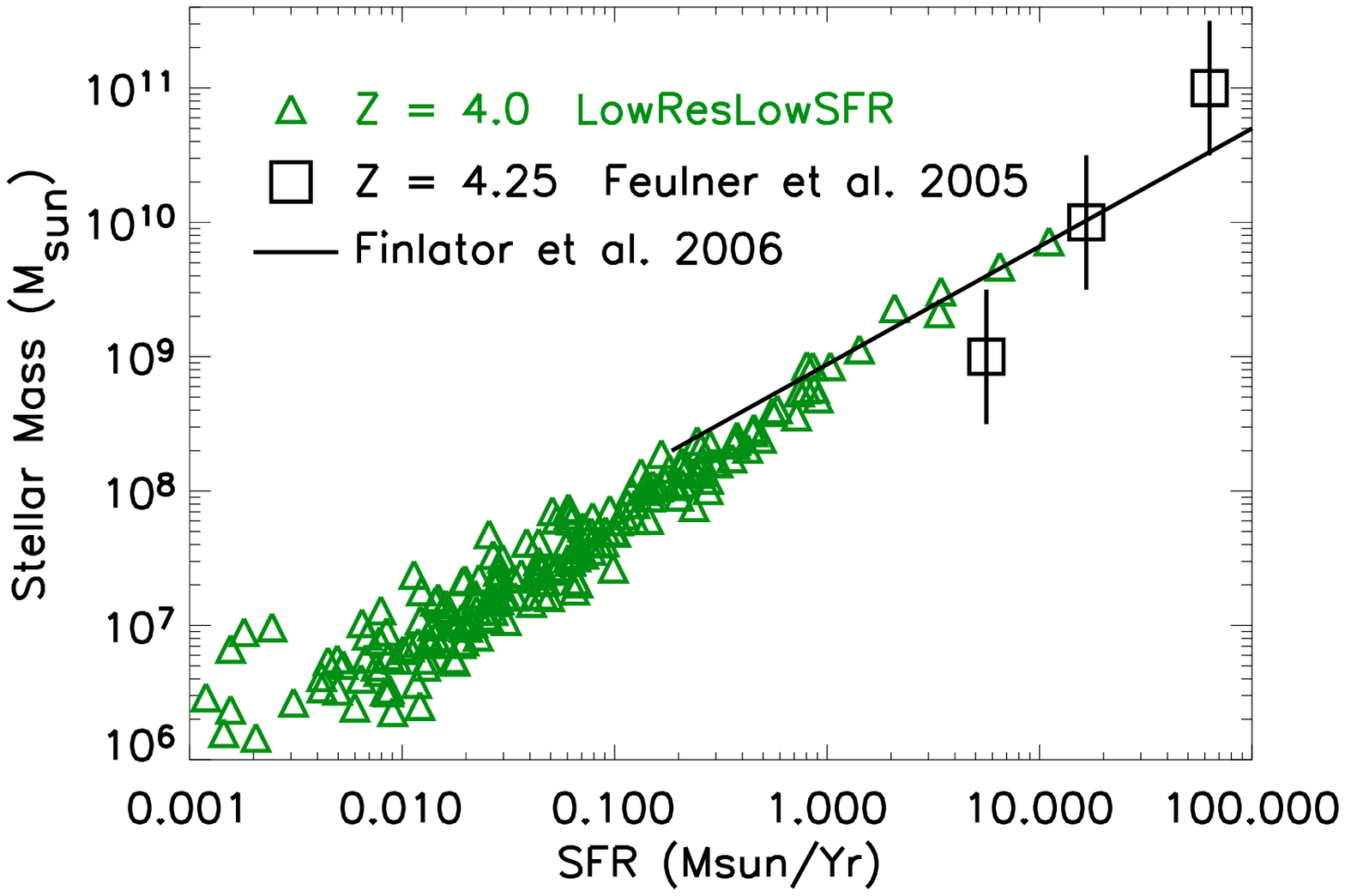}
\plotone{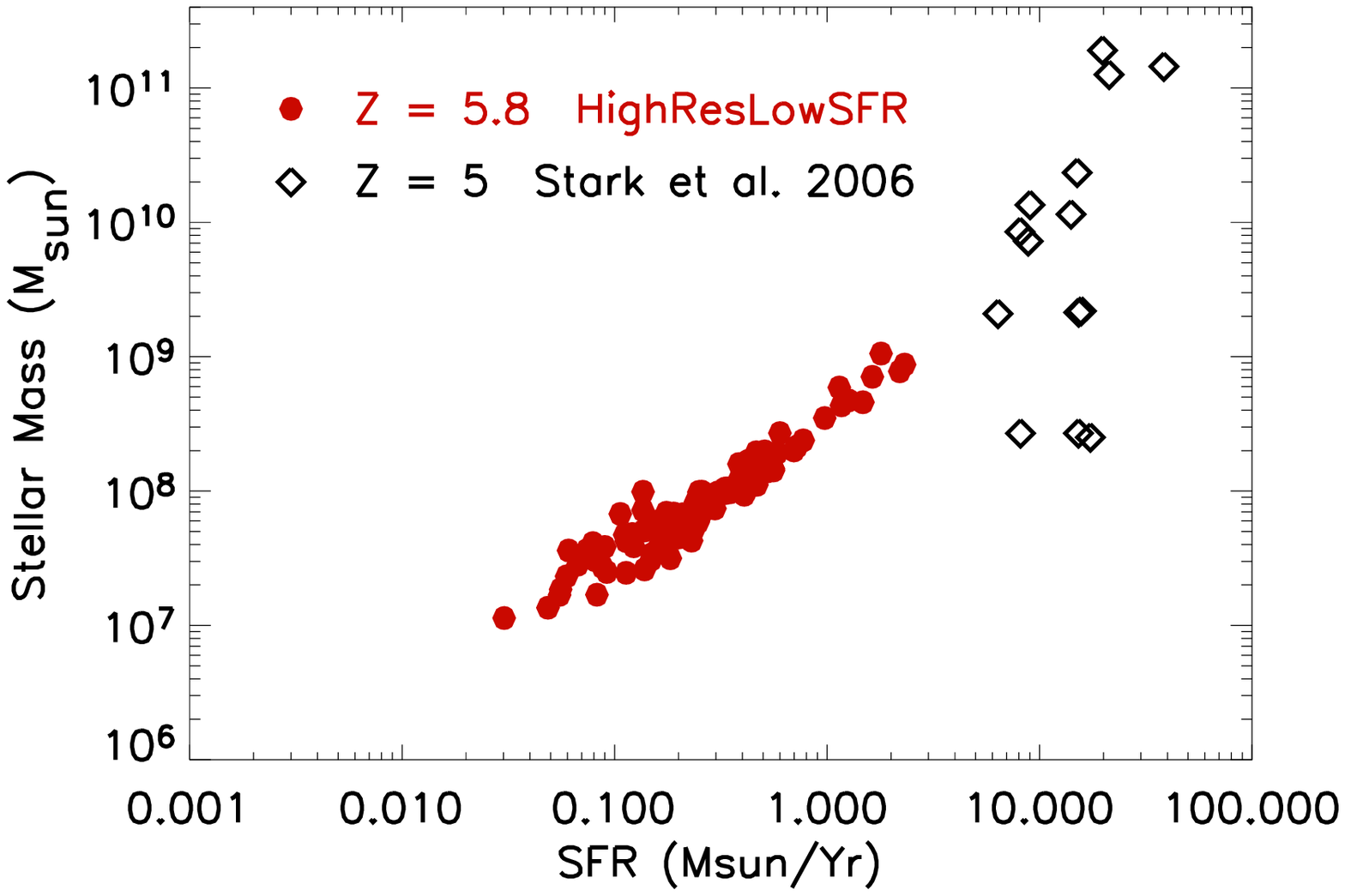}
\caption{Stellar mass as a function of the star formation rate at
  $z=4$ (top panel) and $z=5.8$ (bottom panel) in the observations and
  the simulations (as labeled on the plots).  Simulated galaxies shown
  are those larger than $10^{10}\dim{M}_\odot$.}
\label{stark_fig}
\end{figure}

\begin{figure}[t]
\epsscale{1.0}
\plotone{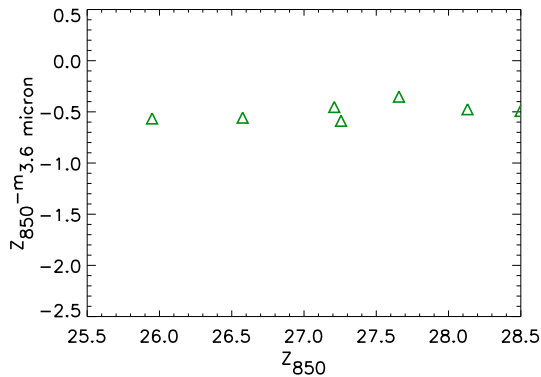}
\plotone{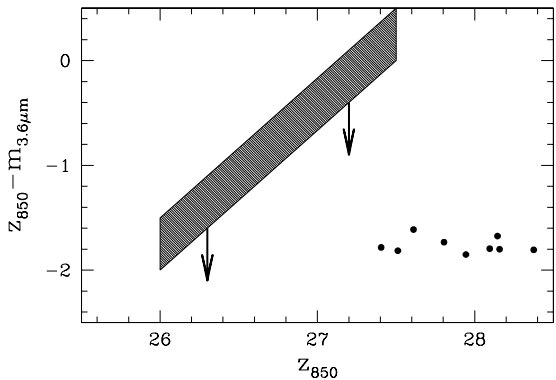}
\caption{The ($\z850-m_{3.6\mu}$) colors vs magnitudes in the
$\z850$ passband for the simulated galaxies at $z=4.0$ (top panel)
and $z=5.8$ (bottom panel).
The shaded band in the bottom panel shows the upper limits 
for the 79 IRAC invisible
galaxies from \citet{yan06}.} 
\label{ir_by_850_fig}
\end{figure}

Recently infrared observations have enabled estimates of the accumulated
stellar mass of high redshift galaxies.  These results place constraints
on prior generations of stars during the epoch of reionization and are
an important test of cosmological simulations.  Much of the observational
data involves galaxies too rare to be present in the simulation box.  Even
so, the trends in the data are interesting.

Figure \ref{stark_fig} shows a power law relation, which is nearly
linear, between the star formation rate and the total stellar mass
content of simulated galaxies at redshifts 4 and 5.8. These results agree
broadly both in slope and magnitude with the hydrodynamic simulations 
of others
\citep{night06,sh03,finlator_06}.  In addition, their data extends 
beyond our highest
points to about $10^{11}\dim{M}_\odot$.
At redshift 4 (top panel) the small overlap with the data of 
\citet{feulner05} is encouraging because the trends are similar.  The
solid line showing good agreement with the trend seen by \citet{finlator_06}
is noteworthy because of the very different techniques used in their
simulations.
The semi-analytic model in \citet{idzi_04}, by contrast, has considerably
more scatter at this redshift.
The spectroscopically verified galaxies of \citet{stark06}
(bottom panel) exhibit a large spread, in contrast to the tight relationship 
in the simulated galaxies.

Contrasting observational results at a redshift of about 6 have been
found by \citet{yan06}.  They find a population of galaxies with very
high stellar masses but low star formation rates.  
These galaxies must have had higher rates of star formation 
in the past since
the current star formation rate is insufficient
to have formed the estimated stellar mass in the time since the Big Bang.
We do not see any counterpart to this population.  However, these
galaxies are presumably very large and thus formed from density
fluctuations too rare to be found in our simulation box; the
effective volume of the \citet{yan06} observations is about 600 times
larger than the volume of our simulation box. 

The observational conversion from magnitude at 3.6 microns to stellar mass
depends upon the unknown star formation histories of the galaxies.
Since we know the star formation history of the simulated galaxies, it
is useful to compute directly the expected observations.  
Fig. \ref{ir_by_850_fig} shows 
$\z850-m_{3.6\mu}$) colors vs magnitude in the $z850$ passband for
our two redshifts.  There is little comparable observational data.
We show in the bottom panel the
IRAC invisible population of \citet{yan06} (Fig. \ref{ir_by_850_fig},
bottom panel).  
These are galaxies, which,
although detectable in the rest frame UV, have too little stellar mass
to be detected in the optical (infrared at this redshift).
Little can be said except that these limits are at least consistent with our
simulation results.

\begin{figure}[t]
\epsscale{1.0}
\plotone{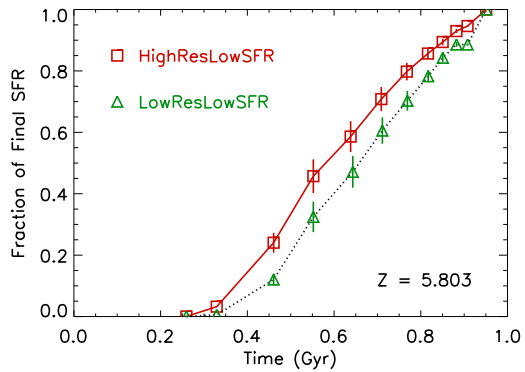}
\caption{The average star formation rate profile for
simulated galaxies at $z=5.8$ with total mass larger 
than $10^{10}\dim{M}_\odot$ 
as a function of time after 
the Big Bang.  The error bars show the standard deviation.}
\label{comb_hist_fig}
\end{figure}

The star formation histories of the simulated galaxies larger than
$10^{10}\dim{M}_\odot$ at
z = 5.8 are remarkably uniform. Figure \ref{comb_hist_fig} shows the
average shape of this history to be a monotonically increasing function.
Some individual galaxies show a decline of star formation rate typically
beginning at 0.8-1.0 Gyr.

\begin{figure}[t]
\epsscale{1.0}
\plotone{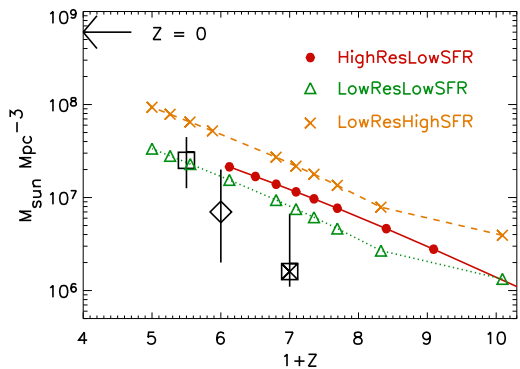}
\caption{The evolution of the total stellar mass density in our three
  simulations (lines with symbols). The square, diamond, and square
  with X 
  show observational results of \citep{drory05}, \citep{stark06}, and
  \citep{yan06} respectively.  Also shown for reference is the present 
  day value from \citet{cole00} (arrow in upper left corner).}
\label{d3_cat_fig}
\end{figure}

Figure \ref{d3_cat_fig} shows the history of the total stellar mass
density of the simulations.  Note that the LowResHighSFR
simulation accumulates about three times as much mass, as one would
expect from the 3 times greater star formation efficiency.
It is difficult to compare these results
to observations since these latter estimates are based on magnitude
limited surveys.  Nevertheless, these results reinforce our decision to
use the lower star formation efficiency. The large arrow, showing
the estimated stellar mass density at $z = 0$ \citep{cole00}, 
provides a ceiling, which should not be crossed.

\section{Conclusions and Discussion}

Despite its pivotal role in the
evolution of the early universe, reionization remains poorly
understood.
As has been previously argued \citep{g00a}, simulations provide 
important dynamical information inherently missing from
analytic and semi-analytic approaches.  This work is part of a 
continuing comparison of detailed
cosmological simulations of reionization with observational data 
as they become available. 

We have shown that numerical simulations of reionization that
reproduce the SDSS data on the evolution of Gunn-Peterson absorption in
spectra of $z\sim6$ quasars can also be made consistent with the observed
galaxy luminosity functions at two different redshifts ($z\sim4$ and
$z\sim6$) {\it simultaneously\/} by adjusting a single parameter:  a
coefficient in the Schmidt law for star formation. The best-fit value of the
coefficient is consistent with the observational estimates from local
galaxies \citep{k98}. In the best-fit case, the star formation
rates per unit galaxy mass are similar to those found by \citet{sh03}
in their simulations using a very different numerical approach. 

The simulations and much of the analysis for this paper
were completed before the third year WMAP results \citep{sea06} 
became available.
The major change for structure formation is a lowering of $\sigma_{8}$.
The expected result is a delay in star formation, but the exact effect, of
course, is not clear without new simulations.  
We have emphasized that the
relative star formation rates at different redshifts are more amenable
to analysis than are the absolute ones.  Our single adjustable
parameter would probably have to be modified for the new cosmology.
\citet{alvarez_06} has estimated the effect of the cosmological 
parameter changes on the collapsed fraction over a range of redshifts.
The ratio of this fraction at $z\sim4$ to that at $z\sim6$ changes by a factor
of about 1.5 between the two cosmologies.  The expected 
direction of the effect
would tend to bring the simulation into better agreement with
observation (see Figure \ref{sfh_fig}).  It should be noted that
the other hydrodynamic simulations to which we have compared our results
have all used the higher value of $\sigma_{8}$.

We are able to obtain agreement with the observed galaxy
luminosity function at $z\sim6$ if we consider all of the luminosity
gravitationally bound to the galaxy and adopt the reddening correction
of \citet{bibf06}. In
addition, the total stellar mass densities at $z\approx5$ to $6$ are
consistent with the latest observational estimates with the
exceptions noted above.

Our results suggest that Lyman Break galaxies observed at $z\approx6$ are 
the brightest ones at that time, with the caveat that 
the finite simulation box size limits
the size of galaxies that we can simulate.

The simulation argues that the amount of star formation we see is
sufficient to reproduce the reionization history imprinted on the
Lyman-$\alpha$ forest.

The luminosity function of the simulation at $z = 5.8$ has 
been compared to the observational
results of \citet{bibf06} because their data overlap the simulation the
most in magnitude.  There seems to be general agreement that 28.5 is the
limiting magnitude at which completeness becomes an issue.  Another
group \citep{bsem04} has taken the conservative approach of confining
the analysis to this limit. They have argued for a lower
star formation rate with the consequence that other reionization sources,
in addition to stars, might be necessary. Our results are not easily
reconciled with these estimates, even when the uncertainty in the free
parameters in the simulation, ie. the star formation efficiency and the
effective escape fraction, are taken into account.

\citet{lpa05} have made an exhaustive study of all galaxies seen to
a limiting magnitude where the redshifts are determined
spectroscopically.  They argue that a significant fraction of galaxies
out to redshift 5 is missed by Lyman Break selection techniques
and that these missed galaxies significantly affect the ionizing
photon budget.
Their magnitude limit is much too low for these galaxies to be
seen in our simulations. 
However, if these conclusions are found to extend
to higher redshifts the simulations will have to be modified to
get a realistic picture of reionization. Larger box sizes, which at
present are prohibitively expensive, may be necessary.  For the
present paper it is important to note that 
their results do support the notion that stars are sufficient for
reionization.

All of our conclusions are limited by the size of the simulation
box. The overall trend seen by \citet{yan06}, in which the the largest
galaxies have seen their zenith in the past, occurs over a much larger
range of galaxy size than we can simulate. Over smaller ranges there
is no clear trend, and thus our results are not inconsistent with
existing data.  However, we might have expected to see at least some
trend in the simulation and we do not.  As discussed by \citet{finlator_06},
the proportionality between star formation rate and stellar mass seen
in hydrodynamic simulations may be an important limitation of these models.
The semi-analytic model of \citet{idzi_04}, although it rests upon more
assumptions, does at least produce more variability.
Simulations with a much higher dynamic range may be needed to
fully account for the population of galaxies contributing to reionization.

Even if it turns out that other reionization sources, not included in
our simulations, are necessary, 
the correspondence of
the simulated galaxies to observations of this period is of
considerable interest since the simulation appears at least 
to reproduce the ionization state
of the universe between redshifts 4 and 6.

\acknowledgments
The authors wish to thank the anonymous referee for helpful
comments that improved the paper.
This work was supported in part by the DOE and the NASA grant NAG
5-10842 at Fermilab, by the NSF grants AST-0134373 and AST-0507596,
and by the National Computational Science Alliance grant AST-020018N,
and utilized IBM P690 arrays at the National Center for Supercomputing
Applications (NCSA) and the San Diego Supercomputer Center (SDSC).
This work was begun while AGH was an associate member of the Center
for Astrophysics and Space Astronomy at the University of Colorado
at Boulder and completed at JILA.  We wish to thank both institutes
and A.\ J.\ S.\ Hamilton of JILA for support.

\bibliography{gayler.rev2,misc,gnedin}

\end{document}